\documentclass[10pt,letterpaper]{article}
\usepackage{opex3}
\usepackage{color}
\usepackage{cite}

\begin{document}

\title{Field and long-term demonstration of a wide area quantum key distribution network}

\author{Shuang Wang,$^{1,2}$ Wei Chen,$^{1,2,5}$ Zhen-Qiang Yin,$^{1,2,6}$ Hong-Wei
Li,$^{1,2,3}$ De-Yong He,$^{1,2}$ Yu-Hu Li,$^{3}$ Zheng
Zhou,$^{1,2}$ Xiao-Tian Song,$^{1,2}$ Fang-Yi Li,$^{1,2}$ Dong
Wang,$^{1,2}$ Hua Chen,$^{1,2}$ Yun-Guang Han,$^{1,2}$ Jing-Zheng
Huang,$^{1,2}$ Jun-Fu Guo,$^{4}$ Peng-Lei Hao,$^{4}$ Mo Li,$^{1,2}$
Chun-Mei Zhang,$^{1,2}$ Dong Liu,$^{1,2}$ Wen-Ye Liang,$^{1,2}$
Chun-Hua Miao,$^{4}$ Ping Wu,$^{4}$ Guang-Can Guo,$^{1,2}$ and
Zheng-Fu Han$^{1,2,7}$}

\address{
$^1$Key Laboratory of Quantum Information, University of
Science and Technology of China, CAS, Hefei 230026, China \\
$^2$Synergetic Innovation Center of Quantum Information \& Quantum
Physics, University of Science and Technology of China, CAS, Hefei 230026, China \\
$^3$Department of Electronic Engineering and Information Science,
University of Science and Technology of China, Hefei 230026,
China;\\
$^4$Anhui Asky Quantum Technology Co.,Ltd. , Wuhu 241002, China \\
$^5$kooky@mail.ustc.edu.cn \\
$^6$yinzheqi@mail.ustc.edu.cn \\
$^7$zfhan@ustc.edu.cn \\
}


\begin{abstract}
A wide area quantum key distribution (QKD) network deployed on
communication infrastructures provided by China Mobile Ltd. is
demonstrated. Three cities and two metropolitan area QKD networks
were linked up to form the Hefei-Chaohu-Wuhu wide area QKD network
with over 150 kilometers coverage area, in which Hefei metropolitan
area QKD network was a typical full-mesh core network to offer
all-to-all interconnections, and Wuhu metropolitan area QKD network
was a representative quantum access network with point-to-multipoint
configuration. The whole wide area QKD network ran for more than
5000 hours, from 21 December 2011 to 19 July 2012, and part of the
network stopped until last December. To adapt to the complex and
volatile field environment, the Faraday-Michelson QKD system with
several stability measures was adopted when we designed QKD devices.
Through standardized design of QKD devices, resolution of symmetry
problem of QKD devices, and seamless switching in dynamic QKD
network, we realized the effective integration between
point-to-point QKD techniques and networking schemes.
\end{abstract}




\section{Introduction}

The principle of a quantum key distribution (QKD) system is to share
secure keys between two remote users. Different from the ways
employing mathematical techniques to avoid eavesdroppers, the
unconditional security of QKD originates from quantum physics, and
has been theoretically proved \cite{gisin2002, scarani}. Since
Muller et al. successfully implemented the first QKD experiment over
23 km installed optical fiber under Lake Geneva \cite{gisin1995},
several groups have demonstrated their QKD experiments in field
environments \cite{hughes1996, townsend, hughes1998, hughes2000,
gisin67km, freespace2002, yuan2005, mo, yuan12, cv}. QKD has moved
far beyond laboratory experiments. Commercial QKD systems are
available from some companies, and moreover, European
Telecommunication Standards Institute has published standards for
QKD to fulfill urgent market needs \cite{standard1, standard2}.

As an important milestone, quantum network, or more precisely QKD
network, was proposed to extend QKD from point-to-point
configuration to multi-user and large-scale scenario
\cite{yuan2007}. Based on the passive beam splitter, Townsend et al.
presented and realized the first QKD network \cite{townsend94,
townsend97}. And since then, researchers have devised and developed
an impressive collection of network architectures for QKD
\cite{circle, WDM, sw03, compare, sw, zhangtao, dianati2008, romain,
wsh10, bus2, mohsen, yuannature, spain}. Combined with increasingly
mature QKD devices, these developments enabled QKD networks to be
deployed over real-world telecommunication networks. Under US
Defense Advanced Research Projects Agency (DARPA), researchers from
BBN Technologies, Boston University and Harvard University built the
world's first QKD network across a metropolitan area -- DARPA
quantum network \cite{darpa1, darpa2}. Then, based on the quantum
router structure \cite{zhangtao}, our group from University of
Science and Technology of China realized the second field QKD
network in the commercial telecommunication fiber network in Beijing
in March 2007 \cite{chenwei}. The SEcure COmmunication based on
Quantum Cryptography (SECOQC) network in Vienna integrated 6
different QKD systems together through trusted repeaters
\cite{SECOQC08, SECOQC09}. Also, ``Q-Government network'' built by
our group was a field trial of application of quantum keys
\cite{xu}. In 2010, the most high-speed QKD network was presented in
Tokyo, where the live video conferencing using one-time-pad (OTP)
encryption was successfully demonstrated \cite{tokyo}. Long-term
performance of QKD networks have also been tested \cite{africa,
idq}, in which SwissQuantum network \cite{idq} ran for more than
one-and-a-half year. And, other QKD networks were also deployed in
the field environment \cite{usa09,pan,pan2010,spain2}.
\begin{figure}[htbp]
\centering\includegraphics[width=10cm]{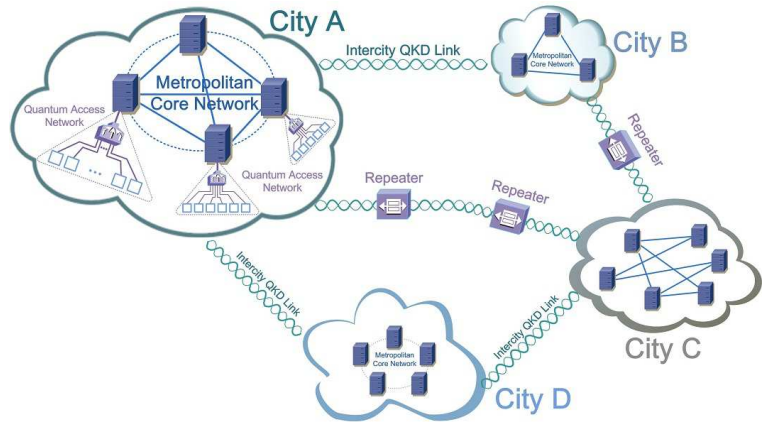}
\caption{Schematic overview of a wide area QKD network.}
\label{fullnetwork}
\end{figure}

In order to furthest extend the applicability of QKD technologies,
QKD networks leverage available fiber networks, and take advantage
of the technologies, components and topologies for conventional
fiber communications \cite{usa09}. Figure \ref{fullnetwork}
illustrates the schematic overview of one QKD network, which is a
wide area QKD network, and for the sake of simple illustration, only
four cities are included. The wide area QKD network is composed of
metropolitan area QKD networks and intercity backbone QKD network
which consists of long-haul intercity QKD links among metropolitan
area QKD networks. The metropolitan area networks usually span
several tens of kilometers \cite{spain}, such as the typical range
of smart cities is 30-80 km \cite{yuanprx}, while distances of
intercity links at least exceed 50 km \cite{intercity}. Although the
transmission distance of QKD systems is already over 250 km
\cite{gisin250,ws260}, quantum techniques including trusted
repeater, quantum repeater \cite{repeater2}, and ground-to-satellite
communication \cite{freespace1} would be used in the intercity link
to extend the distance scale, and multiplex techniques
\cite{duolu,yuan2014} would also be employed to increase the key
rate. The tradeoff between key rates and relay distances is one of
the key issues should be considered before intercity QKD links are
deployed. In each city, the metropolitan area QKD network comprises
metropolitan core networks and quantum access networks. The
metropolitan core network is the core part of the metropolitan area
QKD network, which usually has a mesh configuration to provide
all-to-all interconnections among the network nodes, the
interconnectivity should be taken into account when we plan the core
network. The quantum access network encompasses connections that
extend from one node in the metropolitan core network to multiple
end-users, and provides the last-mile QKD service for subscribers
\cite{yuannature,qnature}. Therefore, quantum access network usually
follows point-to-multipoint topology and covers a few tens of
kilometers \cite{spain}. And, subscribers belonged to different
quantum access networks are connected through the metropolitan core
network.

In this paper, we present a wide area QKD network demonstrated in
the field environment. The network was installed in the Anhui
provincial telecommunication fiber network of China Mobile Ltd.,
with over 150 kilometers coverage area. Three cities and two
metropolitan area QKD networks were linked up by intercity QKD links
to form the wide area QKD network, in which one metropolitan area
QKD network was a typical full-mesh core network based on the QKD
router and switch techniques to offer all-to-all interconnections,
and the other one was a simulative quantum access network using the
1$\times$N switch to realize the point-to-multipoint configuration.
The whole wide area QKD network ran for more than 5000 hours, from
21 December 2011 to 19 July 2012, and part of the network stopped
until last December. To adapt to the complex and volatile network
environment, standardized design of QKD devices and several
networking techniques were tried when the wide area QKD network was
field deployed. And, we also developed and tested two typical
applications of quantum keys on this QKD network.

Compared with previous QKD networks \cite{darpa1, darpa2, chenwei,
SECOQC08, SECOQC09, xu, tokyo, africa, idq, usa09, pan, pan2010,
spain2}, the Hefei-Chaohu-Wuhu QKD network is not only the first
wide area QKD network, but also shows significant improvements on
the QKD networking scheme and technology. In the respect of
networking scheme, for the core QKD network, we designed and
implemented the novel all-to-all scheme based on the combination of
passive and active optical elements, and for the quantum access
network, we proposed and implemented the practical
point-to-multipoint scheme based on the optical switch. Beyond
novelty, these networking schemes make the QKD network more
reliable, flexible and reconfigurable. In the respect of networking
technology, we have successfully resolved two crucial issues -- the
symmetry of QKD devices and seamless switching among different
links. These successes further enhance the competitiveness of QKD
network in the field environment.

\section{Overview of the Hefei-Chaohu-Wuhu wide area QKD network}

\begin{figure}[h]
\centering\includegraphics[width=10cm]{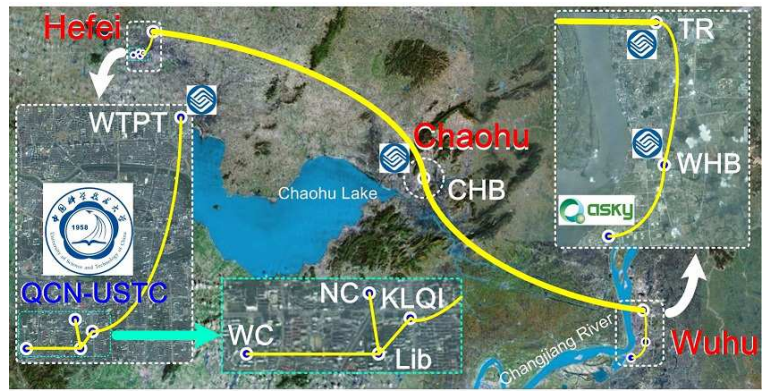}
\caption{Geographic distribution of the Hefei-Chaohu-Wuhu wide area
QKD network, which connects three cities -- Hefei, Chaohu, and
Wuhu.} \label{nodes}
\end{figure}

The Geographic distribution of the wide area QKD network is
presented in Fig. \ref{nodes}, Hefei, Chaohu, and Wuhu are three
connected cities in the network. The Hefei-Chaohu-Wuhu wide area QKD
network consists of three parts: (1) Hefei metropolitan area QKD
network, which has 5 nodes, one node is in Wan-Tong Post and
Telecommunication (WTPT) Co. Ltd., the other 4 nodes are all in the
campuses of University of Science and Technology of China (USTC),
both Library (Lib) and Key Laboratory of Quantum Information (KLQI)
are located in the East Campus, other two nodes respectively locate
in the North Campus (NC) and West Campus (WC), and these 4 nodes
compose the Quantum Campus Network of USTC (QCN-USTC); (2) Wuhu
metropolitan area QKD network, which has 3 nodes that are located in
the Telecom Room (TR) of China Mobile, Wuhu Branch (WHB) of China
Mobile, and Asky Quantum Technology Co. Ltd. (Qasky); (3)
Hefei-Chaohu-Wuhu (HCW) intercity QKD link, which combines Hefei and
Wuhu metropolitan area QKD networks together through the trusted
intermediate node lies in the Chaohu Branch (CHB) of China Mobile. 6
nodes are located in the north of Changjiang river, and the other 3
nodes are located in the south of Changjiang river, hence the HCW
wide area QKD network is the first QKD network across the Changjiang
river.

There are 8 optical fiber links among 9 nodes in the wide area QKD
network, as the yellow lines in Fig. \ref{nodes}. All the 8
optical fiber links are characterized as table \ref{link}. The total
length of the installed fiber is almost 200 km, of which the
intercity fiber length exceeds 150 km.
\begin{table}[htb]
\centering\caption{Characteristics of fiber links in the
Hefei-Chaohu-Wuhu wide area QKD network.} \label{link}
\begin{tabular}{c c c c}
\hline \hline                        &\textbf{Optical fiber link}   &\textbf{Length of fiber}(km) &\textbf{Optical loss}(dB)     \\
\hline \textbf{Hefei metro network}  &WTPT--USCT(KLQI)              &16.9                         &$-$6.1                                       \\
                                     &KLQI--Lib                     &0.9                          &$-$1.2                                          \\
QCN--USTC                            &NC--Lib                       &1.2                          &$-$0.6                                       \\
                                     &WC--Lib                       &1.9                          &$-$0.5                                                  \\
\hline
\textbf{HCW-intercity link}          &Hefei--Chaohu                 &85.1                         &$-$18.4                                             \\
                                     &Chaohu--Wuhu                  &69.7                         &$-$14.1                                              \\
\hline
\textbf{Wuhu metro network}          &TR--WHB                        &14.3                         &$-$5.0                                             \\
                                     &WHB--Qasky                     &9.0                          &$-$7.1                                         \\
\hline\hline
\end{tabular}
\end{table}

\begin{figure}[htbp]
\centering\includegraphics[width=10cm]{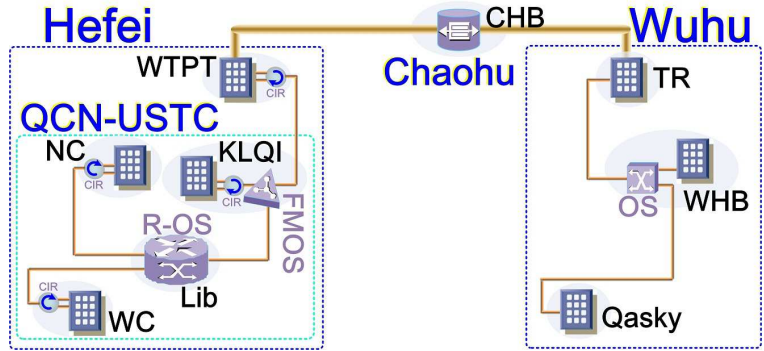} \caption{Topology of
the wide area QKD network. CIR: optical circulator, OS: optical
switch, FMOS: full-mesh optical switch, R-OS stands for the
combination of the QKD router and optical switch.} \label{top}
\end{figure}

Figure \ref{top} shows the topology of the whole wide area QKD
network. Two kinds of networking schemes \cite{tokyo,idq} were
adopted in the wide area QKD network. One is based on trusted
intermediate nodes, there were three trusted intermediate nodes in
the three parts of the whole QKD network respectively: WTPT in the
Hefei metropolitan area network, TR in the Wuhu metropolitan area
network, and CHB in the HCW-intercity link. The other one is based
on additional optical components, in which the fiber was shared
among multiple nodes, both Hefei and Wuhu metropolitan area QKD
networks were realized in this way. The Hefei metropolitan area
network is full-mesh, each network node has direct link with all the
other nodes in this network. And, the Wuhu metropolitan area QKD
network is a time division multiplexing type, only one
point-to-point (P2P) QKD link exists at each time. Here, the
full-mesh Hefei metropolitan area QKD network is one typical
metropolitan core network to offer all-to-all interconnections,
while the Wuhu metropolitan area QKD network is used to simulate a
quantum access network.

\section{P2P QKD devices in the wide area QKD network}

In the HCW wide area QKD network, a total of 13 QKD devices were
employed to support the two metropolitan area QKD networks and
intercity QKD links. To manufacture and maintain these QKD devices,
it would require a large amount of work. Therefore, standardized
design of QKD devices was tried before deployment of the wide area
QKD network, especially for the symmetry among QKD devices which
will be discussed later. All P2P QKD devices adopted the
Faraday-Michelson interferometer (FMI) system \cite{mo} to implement
phase coding BB84 protocol \cite{bb84} with the decoy state method
\cite{decoy1, decoy2, decoy3}. And, all parts of each QKD devices
were housed in a standardized 3.5U rack mount case
(15.6$\times$44$\times$40, H$\times$W$\times$D, cm), in which the
optics, associated electronics, the single photon detector, and one
computer with a displayer and a keyboard were all included. We
designed two types of P2P QKD devices: one is the divided type, in
which the QKD-Transmitter and the QKD-Receiver were separated in two
cases respectively; the other was the integral type, named
QKD-Transceiver, in which the QKD-Transmitter and the QKD-Receiver
were integrated together in one case with an optical switch.

\subsection{The divided type -- QKD-Transmitter and QKD-Receiver}

\begin{figure}[htbp]
\centering\includegraphics[width=9cm]{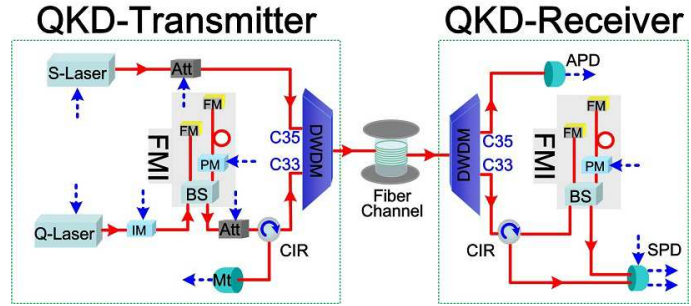} \caption{Schematic
of the divided QKD device.} \label{system}
\end{figure}

Figure \ref{system} outlines the divided type QKD device. Both sync
and quantum signals are transmitted through the same fiber channel.

In the QKD-Transmitter, the sync signal of 1549.32 nm (channel C33
of the ITU grid) wavelength is produced by the sync laser (S-Laser),
and attenuated before entering the dense wavelength division
multiplexing (DWDM). The quantum laser (Q-Laser) generates the
1550.92 nm (channel C35 of the ITU grid) pulse train with 20 MHz
repetition rate and 600 ps pulse width. These quantum pulses are
first prepared in decoy states by an intensity modulator (IM), then
phase-encoded through the phase modulator (PM) in the
Faraday-Michelson interferometer (FMI), and finally attenuated to
single-photon level using an attenuator (Att). The 3-port optical
circulator (CIR) and monitor (Mt) are added to prevent and detect
possible Trojan-horse photons from the channel.

In the QKD-Receiver, DWDM is used to demultiplex the received sync
and quantum pulses. The sync pulses is detected by a normal
avalanche photodiode (APD) to keep the QKD-Receiver and
QKD-Transmitter in synchronicity. After passing through CIR,
phase-encoded quantum pulses are first phase-decoded by PM in the
paired FMI, and then detected by the single photon detector (SPD).
The InGaAs/InP APDs belonged to SPD is cooled down to
-50$^{\circ}$C, and works in Geiger mode with a gate width of less
than 1 ns.

\subsection{The integral type -- QKD-Transceiver}

\begin{figure}[htbp]
\centering\includegraphics[width=8cm]{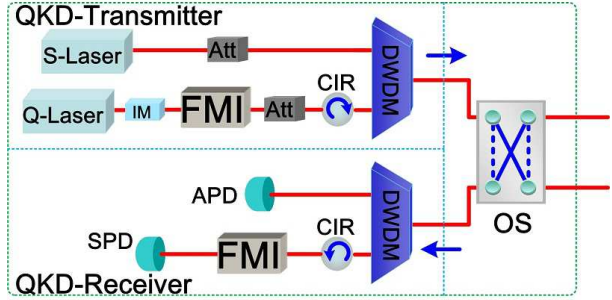} \caption{Schematic of
the QKD-Transceiver.} \label{tr}
\end{figure}

Figure \ref{tr} shows the integral type QKD system that is named
QKD-Transceiver. It consists of a QKD-Transmitter, a $2\times2$
optical switch (OS), and a QKD-Receiver. The $2\times2$ OS has two
sates: in the ``cross'' state (solid lines of OS in Figure \ref{tr})
the QKD-Transmitter part communicates with remote QKD-Receiver, and
the QKD-Receiver part communicates with remote QKD-Transmitter
respectively to share secure keys; in the ``bar'' state (dash lines
of OS in Figure \ref{tr}) the QKD-Transmitter and the QKD-Receiver
parts communicate with each other to calibrate parameters such as
half-wave voltages and delay times, and the QKD-Transceiver is
skipped in the network. Incidentally, the QKD-Transceiver might be a
good form of the trusted repeater.

\subsection{The security consideration of QKD devices}

Practical security is a key consideration when QKD devices are built
up. Many quantum hacking strategies based on imperfections of
practical devices have been proposed, and some of them have been
experimentally implemented successfully. For P2P QKD devices in HCW
wide area QKD network, some proposed hacking strategies were
considered, and corresponding counter-measures have been added into
the QKD system as follows.

The decoy state method \cite{decoy1, decoy2, decoy3} was combined
with BB84 protocol in the QKD device to exclude the PNS attack
\cite{scarani}. As a benefit, this method can dramatically increase
the secure key rate. The decoy state method is implemented as
follows: IM in the QKD-Transmitter is used to create the signal
pulses of 0.65 (or 0.7) photons per pulse and the decoy pulses of
0.1 photons per pulse, the vacuum pulses are generated by not
triggering Q-laser, and the ratio among the signal pulses, decoy
pulses and vacuum pulses is 14:1:1.

Four phases ($\{0, \pi\}, \{\frac{\pi}{2}, \frac{3\pi}{2}\}$)
modulation scheme was implemented in the QKD-Receiver to neutralize
the fake state attack \cite{fake} and time-shift attack
\cite{timeshift} which utilize the detection efficiency mismatch in
the time domain. To restrain costs, only one InGaAs/InP APD was used
in SPD for most QKD-Receivers, in which the four phases modulation
is necessary.

Real-time physical random number generators (RNG) provided QKD
devices with random bits. Quantum RNG is a critical component of QKD
devices, but the commercial products such as Quantis from id
Quantique is still difficult to meet the speed demand for QKD
devices. As a compromise between speed and security, 20 Mbps
physical RNG (WNG-8 from Beijing Hongsi Electronic Tech. Co. Ltd.
\cite{hs}) based on thermal noise was adopted in our QKD devices. 6
pieces of such RNG were used in the QKD-Transmitter for random decoy
state preparation (4$\times$20 Mbps) and phase modulation
(2$\times$20 Mbps), and 2 pieces were used in the QKD-Receiver for
phase demodulation.

The SPD component of QKD devices was transformed to avoid being
controlled by bright light \cite{bright,avoid,avoid2}, and
monitoring technologies was employed to detect bight light that
sneaks into the QKD device. In addition to monitoring the
photocurrent, the normal APD for sync signal detection was also an
auxiliary monitor, because the isolation between C33 and C35 of DWDM
is limited, usually -35 dB. And, for QKD-Receivers with only one APD
in SPD, the output of FMI without connecting SPD was also monitored
by normal APD.

Although practical security of QKD devices has been considered in
this network, techniques such as device-independent QKD \cite{di} or
measurement-device-independent QKD \cite{mdi} might be used in the
future to further enhance the practical security.

\section{Detailed layout of the HCW wide area QKD network}

The QKD network is not a simple combination of QKD devices, but an
effective integration between QKD devices and networking schemes (or
techniques). In the HCW wide area QKD network, Hefei metropolitan
part was based on QKD router and switch techniques to realize the
full-mesh structure with 6 QKD devices, Wuhu metropolitan part used
the 1$\times$N switch to realize the point-to-multipoint
configuration with N+1 QKD devices, and HCW-intercity link connected
these two metropolitan part together through the trusted repeater
technology. In particular, multiplexing are very useful techniques
that share optical fiber channels and QKD devices to realize
reliable, cost-competitive and reconfigurable QKD networks. We have
resolved two key problems of multiplexing techniques - the symmetry
of QKD devices and seamless switching in the HCW wide area QKD
network.

\subsection{Full-mesh Hefei metropolitan area QKD network}

\begin{figure}[htbp]
\centering\includegraphics[width=10cm]{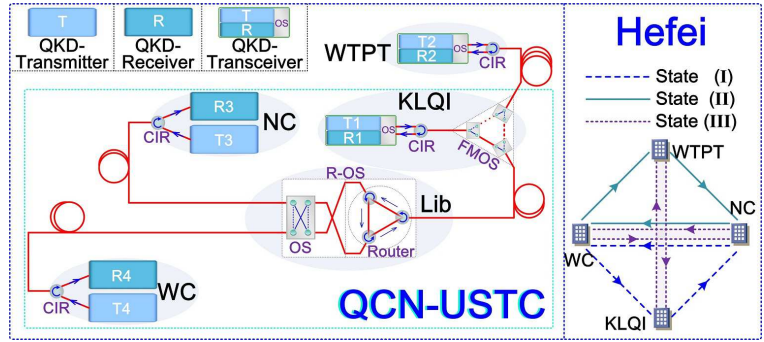} \caption{Structure
of full-mesh Hefei metropolitan area QKD network.} \label{hefei}
\end{figure}

The Hefei metropolitan area QKD network is a full-mesh network with
4 QKD nodes as shown in Fig. \ref{hefei}. Each QKD node in the
network had direct physical links to all the other nodes, although
there were only 4 installed fiber channels. To share the limited
fiber channel resource, we adopted QKD router and switch techniques,
specifically, the wavelength-saving real-time full-mesh (RTFM) QKD
router \cite{wsh10}, and the time division multiplexing full-mesh
optical switch (FMOS). Here, the router and switch served as optical
components that were responsible for dynamically routing and
blocking in the QKD network.

The structure of wavelength-saving RTFM router was propose in
reference \cite{wsh10}. With $N$ wavelengths, the wavelength-saving
RTFM router can support a RTFM QKD network with $2N+1$ QKD nodes, in
which every two nodes share secure keys directly at the same time,
and each node only has one fiber connecting with the router. Here, 3
CIRs were used to compose the simplest form of this router (see the
router part of the R-OS in Fig. \ref{hefei}). Using single
wavelength, this router can support a RTFM QKD network with 3 nodes,
and every 2 nodes can share secure keys directly at the same time.

The structure of time division multiplexing FMOS is very similar
with normal RTFM QKD router consisting of WDMs \cite{zhangtao}, each
N wavelengths WDM is replace by a $1\times N$ OS. Using FMOS with
$N+1$ ports, no more than $(N+1)/2$ pair nodes can share secure keys
directly at the same time, but every two nodes can distribute secure
keys with each other by the time division multiplexing method. Here,
three $1\times2$ optical switches were used to compose the simplest
form of time division multiplexing FMOS (see the FMOS in Fig.
\ref{hefei}). The FMOS can also support a full-mesh QKD network with
3 nodes, but only two nodes are connected at every time.

By combing the QKD router and FMOS as above, the full-mesh Hefei
metropolitan area QKD network had three states (as shown in the
right part of Fig. \ref{hefei}) corresponding to the three
connection statuses of FMOS respectively.

\emph{State (\uppercase\expandafter{\romannumeral 1}) -- KLQI, NC
and WC comprised the RTFM QKD network}: the $2\times2$ optical
switch was in the ``cross'' state, and the QKD-Transceiver at WTPT
worked in the self-calibration mode, the quantum signals were from
T1 to R3, from T3 to R4, from T4 to R1, and from T2 to R2,
respectively.

\emph{State (\uppercase\expandafter{\romannumeral 2}) -- WTPT, NC
and WC comprised the RTFM QKD network}: the $2\times2$ optical
switch was still in the ``cross'' state, and the QKD-Transceiver at
KLQI worked in the self-calibration mode, the quantum signals were
from T2 to R3, from T3 to R4, from T4 to R2, and from T1 to R1,
respectively.

\emph{State (\uppercase\expandafter{\romannumeral 3}) -- KLQI and
WTPT, NC and WC had duplex QKD links between each other}: the
$2\times2$ optical switch was in the ``bar'' state, and the quantum
signals were from T1 to R2, from T2 to R1, from T3 to R4, and from
T4 to R3, respectively.

Through controlling FMOS in the KLQI and $2\times2$ OS in the Lib,
Hefei metropolitan area QKD network was in one of aforesaid tree
states. There were 8 different direct QKD links in this QKD network,
it was enough to support a full-mesh metropolitan core network with
4 QKD nodes.

\subsection{Point-to-multipoint Wuhu metropolitan area QKD network}

\begin{figure}[htbp]
\centering\includegraphics[width=8cm]{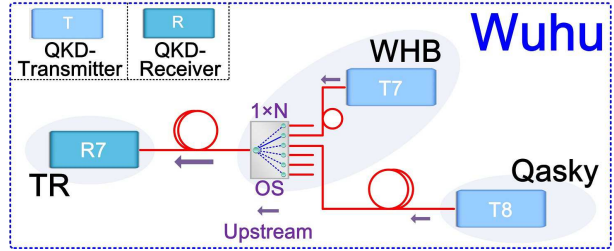} \caption{Structure
of point-to-multipoint Wuhu metropolitan area QKD network.}
\label{wuhu}
\end{figure}

Wuhu metropolitan area QKD network is a quantum access network with
three QKD nodes which are connected by a simple 1$\times$2 optical
switch with 1 output and 2 input ports. It is a typical
point-to-multipoint access network with time division multiplexing,
only one end node accesses QKD link at each time. As shown in Fig.
\ref{wuhu}, two QKD-Transmitters T7 and T8, which were respectively
placed in WHB and Qasky, transmitted signals to the 2 input ports of
the optical switch at different times, and one QKD-Receiver R7 which
located in TR received signal from the output port of the optical
switch. At each time, QKD-Receiver R7 only communicated with the
QKD-Transmitters T7 or T8.

Replacing the 1$\times$2 optical switch with 1$\times$N one, this
quantum access network could support N access nodes, with N
QKD-Transmitters and only one QKD-Receiver. Since the QKD-Receiver
requires SPDs, which are often expensive, difficult to
operate\cite{yuannature}, and bulky, the recommendable way is the
upstream implementation, one QKD-Receiver receives signals from
multiple QKD-Transmitters in the point-to-multipoint configuration.
Beam splitter, WDM and optical switch are three optical components
conveniently providing point-to-multipoint connections. Beam
splitter has a high insertion loss about $-10\lg N$ dB, WDM needs
different laser sources, therefore, optical switch might be a
compromise between beam splitter and WDM, but it requires a
auxiliary system to offer power supply and remote control.

\subsection{HCW-intercity QKD link}

HCW-intercity QKD link connects Hefei and Wuhu cites, the distance
and channel loss between which exceed 150 km and -32 dB
respectively. In order to obtain practicable key rate, Chaohu city,
which is located between Hefei and Wuhu, was chosen as the trusted
intermediate node. Through the HCW-intercity QKD link, any node in
Hefei metropolitan area QKD network could share share secure keys
with any node in Wuhu metropolitan area QKD network using the
hop-by-hop fashion \cite{romain}. The optical loss of fiber channel
between Hefei and Chaohu was -18.4 dB, but the loss between Chaohu
and Wuhu was only -14.1 dB. Therefore, the best-matched
QKD-Transmitter T5 and QKD-Receiver R5 were picked out to share
secure keys between node WTPT in Hefei and node CHB in Chaohu, and
the best two APDs were selected to compose SPD in R5. While, the
second best-matched QKD-Transmitter T6 and QKD-Receiver R6 were used
to share secure keys between node CHB in Chaohu and node TR in Wuhu,
but only one APD was used in R6, the same as all the other
QKD-Receivers. The Chaohu to Wuhu link is the first QKD link across
the Changjiang river.

\subsection{Symmetry of QKD devices}

Symmetry of QKD devices is a pivotal issue for QKD networks,
especially for the multiplexing type, in which one QKD device needs
to communicate with multiple QKD devices. As to this specific wide
area QKD network, in the Hefei metropolitan part, the
QKD-Transmitter T1 needs to communicate with three different
QKD-Receivers R3, R1, and R2 in the aforesaid three states, R3 needs
to communicate with T1, T2, and T4, etc.; in the Wuhu metropolitan
part, the same QKD-Receiver R7 has to communicate with two different
QKD-Transmitters T7 and T8 in the time division multiplexing way.
For these one-way phase coding QKD devices, the core symmetry
problem is the matching requirements of every unbalanced
interferometers - FMIs. Therefore, we have developed the method to
fabricate FMIs, of which properties were almost the same. And,
symmetry of QKD devices is finally characterized by measuring the
quantum bit error rate (QBER) in back-to-back transmission. The
measured results are given in table \ref{sym}. From this table,
QBERs of QKD-Transmitters (T1, T2, T3, T4) and QKD-Receivers (R1,
R2, R3, R4) are all below 1.20\%, and we may draw a conclusion that
the symmetry problem of QKD devices has been largely resolved. If
one QKD device in the network breaks, we only need to replace the
broken one rather than paired devices, and if one new QKD node want
to join the quantum access network, we just need one
QKD-Transmitter, it makes the QKD technology more cost-competitive
in the network respect.

\begin{table}[htb]
\centering\caption{Symmetry characteristic of QKD devices}
\label{sym}
\begin{tabular}{c c c c c}
\hline \hline \textbf{QBER}                       &\textbf{R1}            &\textbf{R2}            &\textbf{R3}               &\textbf{R4} \\
\hline \textbf{T1}                                &\textcolor[rgb]{0.00,0.00,1.00}{0.97\%}                 &\textcolor[rgb]{0.00,0.00,1.00}{0.85\%}                 &\textcolor[rgb]{0.00,0.00,1.00}{0.70\%}                    &1.12\%       \\
\textbf{T2}                                       &\textcolor[rgb]{0.00,0.00,1.00}{0.48\%}                 &\textcolor[rgb]{0.00,0.00,1.00}{1.02\%}                 &\textcolor[rgb]{0.00,0.00,1.00}{0.53\%}                    &1.07\%       \\
\textbf{T3}                                       &0.86\%                 &0.91\%                 &1.09\%                    &\textcolor[rgb]{0.00,0.00,1.00}{0.78\%}        \\
\textbf{T4}                                       &\textcolor[rgb]{0.00,0.00,1.00}{0.56\%}                 &\textcolor[rgb]{0.00,0.00,1.00}{0.62\%}                 &\textcolor[rgb]{0.00,0.00,1.00}{0.49\%}                    &0.67\%        \\
\hline\hline
\end{tabular}
\end{table}

\subsection{Seamless switching in QKD networks}

Seamless switching is helpful to improve performances of dynamic QKD
networks that have several states to switch back and forth, for
example, QKD-Receiver R7 in Wuhu metropolitan area QKD network has
to switch between two states to dynamically distribute keys with
QKD-Transmitters T7 or T8. In QKD networks, the switching process
includes not only changing the status of optical switching
equipments, but also establishing a new QKD link or several new QKD
links. Although the switching is dynamic, the states before and
after switching are deterministic, which means the QKD device knows
in advance who it will communicate with. Therefore, QKD devices
could stored configuration parameters that need to communicate with
other possible devices when the dynamic QKD network is deployed,
then establishment new QKD links are simplified to retrieve
configuration parameters. And as mentioned above, the phase coding
system based on unbalanced FMI was adopted, and the sync and quantum
signals were multiplexed in the same fiber with 1.6 nm wavelength
interval, these measures make the stored configuration parameters
are very stable for establishing new QKD links. In Hefei and Wuhu
metropolitan area QKD networks, establishing new QKD links were
proceeded simultaneously with changing the status of optical
switches. For this reason, the switching process was very quick, and
could be regarded as seamless. We designed two seamless switching
mode here, the preemptive mode and automatic mode. If there were
some special requirements, the preemptive switching mode worked, and
the QKD network switched to and kept in the required state. At other
times, the QKD network was in the automatic switching mode, and
automatically switched to different states at set intervals to
periodically updates secure keys.

\section{Long-term performance with the field environment}

The field test assesses both performance and reliability of QKD
networks in the actual operation environment. The field environments
of the HCW wide area QKD network were relatively complicated: the
optical fiber channels included the metropolitan area optical
networks and intercity links, and operation environments of QKD
nodes involved the telecom room, the laboratory, the normal room,
and even the makeshift kitchen. Further, this whole QKD network ran
for more than 5000 hours, from 21 December 2011 to 19 July 2012, and
the the QCN-USTC part stopped until last December. For this long
term, the field environment has changed a lot. For QKD networks,
reliability is as essential as overall secure key rate. This effects
not only the design of QKD devices, but also the topology of the QKD
network. Therefore, the phase coding system based on FMI was adopted
against fluctuations on the fiber channel, and the complex full-mesh
topology was implemented in Hefei metropolitan core network to offer
high flexibility and interconnectivity.

\subsection{The field fiber channels and measures of crosstalk reduction}

In the HCW wide area QKD network, most of the installed optical
fiber channels were commercial fibers provided by China Mobile Ltd.,
and all fibers were standard telecom fiber (ITU-T G.652). From
measurement results of the optical time domain reflector (Yokogawa,
AQ7275), the fiber channels in intercity backbone links were very
good - with no reflective events and 0.21 dB/km average loss
coefficient, while, the fiber channels in metropolitan area were not
good - with many reflective events and 0.46 dB/km average loss
coefficient. In metropolitan area optical network, many splicing
points and connectors combined numerous short fibers together, and
reflective events from some splicing points and connectors with bad
return loss made signals (including quantum and sync signals) in the
channel influence each other, especially when several QKD links
shared the same channel at the same time.

We improved the crosstalk suppression at the device level. To
minimize the reflective noise on the quantum signals, nonadjacent
channels of DWDM were chosen to transmit quantum and sync signals
respectively, and between which time delay module was added. The
field fiber channels were parallel with other commercial fibers in
which high power optical streams were transmitted in the same
multi-core cable, and these parallel fibers would leak photons into
quantum channels \cite{leak}, which made field fiber channels not
dark any more. Different from forgoing reflective noise that could
be considered as point events, the leakage photons would be
continuous noise in the time and spectrum domain. However, this
leakage noise could be greatly reduced by the DWDM and the time gate
of the SPD, which could be regarded as a spectral filter and a
temporal bandpass respectively.

\subsection{The operation environments of QKD devices}

\begin{figure}[htbp]
\centering\includegraphics[width=12cm]{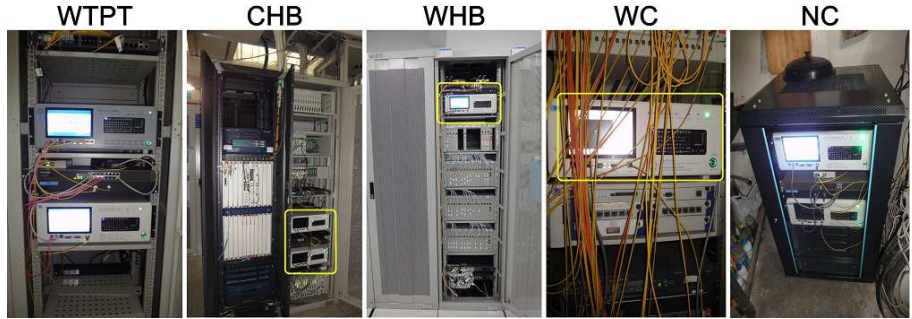}
\caption{Photographs of some nodes, including the surrounding
environments and QKD devices.} \label{photo}
\end{figure}

The performance of QKD devices is not only affected by the fiber
channel, but also by the operation environments, which include
environment temperature, humidity, pressure, dust, and ambient
vibration. Figure \ref{photo} shows photographs of five QKD nodes in
HCW wide area QKD network, WTPT and CHB nodes in the intercity QKD
link, WHB node in Wuhu metropolitan area QKD network, WC and NC
nodes in the QCN-USTC part of Hefei metropolitan QKD network. The
operation environments of QKD devices can be divided into four
types: the telecom room, the simple switching room, the office room,
and the makeshift room.

Nodes WTPT, CHB, TR, and WHB were in the telecom rooms provided by
China Mobile Ltd.. These telecom rooms have precision air
conditioning systems, which control the temperature, humidity, and
dust within tight tolerances. And, the telecom room has relatively
strong physical security, anyone who wants to enter the room needs
to be authenticated, and video cameras always monitor the room.
These good environmental conditions were beneficial to long-term
stable operation of QKD systems. However, the vibration and acoustic
noise (e.g. nearby fans and humming equipment) in the telecom room
are unneglectable and sustained, which requires the QKD devices able
to resist a certain intensity of vibration.

Nodes WC and Lib were in the simple switching rooms of USTC. These
simple rooms are relatively separate and only equipped with a
household air conditioning for cooling. The temperature and humidity
of these simple rooms are controlled within loose tolerances, and
there is a lot of dust and also vibration from fans of nearby
equipments.

Nodes KLQI and Qasky were in the office rooms, or normal
laboratories rooms. These offices have a central air conditioning
system to make people comfortable, so the temperature and humidity
are 18 \textordmasculine C $\sim$ 28 \textordmasculine C and 30\%
$\sim$ 60\% respectively when staff are in the office, but the
central air conditioning system and lights are shut down when staff
are off duty. There are twice temperature variation process on
weekdays at least, especially in the winter and summer. In addition,
vibration from staff activity should also be considered.

Nodes NC was in the makeshift kitchen of the doorkeepers, which was
the harshest operation environment in the network. There are a gas
cooker, a small exhaust fan, and other things to cook Chinese food.
The kitchen become a hot, noisy, and smoky place when doorkeepers
make the lunch and supper. Therefore the QKD-Transmitter T3 and
QKD-Receiver R3 were put into a telecom cabinet to keep away from
the lampblack. Since there is no air conditioning, the winter
temperature falls to minus 5 \textordmasculine C, while the summer
temperature gets up to 32 \textordmasculine C, and even higher when
cooking lunch in summer, which requires large operating temperature
range for QKD devices, especially for the SPD part.

\subsection{The influence from weather and measures for QKD}

Weather is a crucial factor for the long-term operation of a wide
area QKD network. During the operation term, weather would vary
greatly. The influence from weather included on the QKD devices and
the fiber channels. Changes of fiber channel properties would
indirectly affect the stability and performance of QKD devices, and
some measures were adopted to automatically compensate influence
from the weather.

The variation of the weather would directly affect the operation
status of QKD devices, especially for those in the nodes without air
conditioning systems. Although several temperature control modules
should be adopted to improve the stability of QKD devices, the
ambient temperature and humidity are still important factors. In
this network, the greatest challenge is to keep  the working
temperature of the SPD at -50 \textordmasculine C. Most SPDs could
work normally, only the SPD of QKD-Receiver R3 in NC node stopped
working on the afternoon of June 8th, and resumed in the evening.
Although the limited operating temperature of the SPD was about 35
\textordmasculine C, the ambient temperature and humidity, and also
the cooking activity made the environment very tough for the cooling
module belonged to the SPD part. In order to keep the working sate
of R3 in the whole summer, the working temperature of its SPD part
was changed from -50 \textordmasculine C to -40 \textordmasculine C
(the same as researchers did in the SwissQuantum QKD network
\cite{idq}). After being reset the working temperature, the
detection efficiency of the SPD was adjusted the same as before, but
the dark count rate doubled.

The change of the weather would vary the birefringence, optical path
length (or time delay), and loss of fiber channels, which are the
main differences from the laboratory experiments. Moreover, these
variations accumulate with distance. The operation stability of QKD
devices would be greatly affected by variations of birefringence and
optical path length of fiber channels. Instead of adding
stabilization mechanism, we adopted two measures to automatically
compensate these two types of variations of channels respectively.
By using phase coding system based on FMI, QKD devices were
insensitive to the polarization disturbance of channels \cite{mo}.
The other measure was multiplexing quantum and sync signals in the
same fiber with only 1.6 nm wavelength interval to compensate
changes in optical path length, which led to drift of
synchronization between the QKD-Transmitter and QKD-Receiver. Even
with 50 \textordmasculine C temperature change, the drift of
synchronization in Hefei-Chaohu fiber channel was much less than 1
ps, which was negligible. The results from theoretical analysis and
experiments show the effectiveness of these two measures. However,
variations of loss of fiber channels are unavoidable, and their
influences could not be compensated since intensity of each quantum
signal from QKD-Transmitters is fixed.

\subsection{Other influence factors}

In addition to the weather, there were also some external factors
influencing the performance of the long-term QKD network: (1) Fiber
cable breaking by the road construction. One road near Node Qasky
was under construction in the meantime. The fiber cable from Node
WHB to Node Qasky was broken three times during the network
operation phase, which made Node Qasky to be isolated from the whole
network. (2) Power outage. All QKD devices in the wide area QKD
network could operate from the single-phase AC power source.
However, the stable power source in the telecom room is 48 V DC
power. At the beginning, the lighting AC source (220V, 50Hz) in the
wall was used in Node WTPT. Only ten days after running the whole
QKD network, there was a power outage, which made the wide area
network to be split into two independent metropolitan network.
Considering the stability of the power source, one DC-AC inverter
was used to transfer the stable 48 V DC power to 220 V, 50 Hz AC
power in Node WTPT. (3) Computer crash. (4) Halt of the controlling
system.

\subsection{Long-term performance of QKD links}

After obtaining characteristics of the eight optical fiber links in
the wide area network, we first built the whole QKD network in the
laboratory. The test run lasted for one week, and then all devices
were carried to the corresponding field Nodes. The whole wide area
QKD network was completed on 20 December 2011. Starting from the
next day and stopping on 19 July 2012, the field test of the whole
wide area QKD network lasted more than 5000 hours, and the test of
the QCN-USTC part stopped until last December.

\begin{figure}[htbp]
\centering\includegraphics[width=10cm]{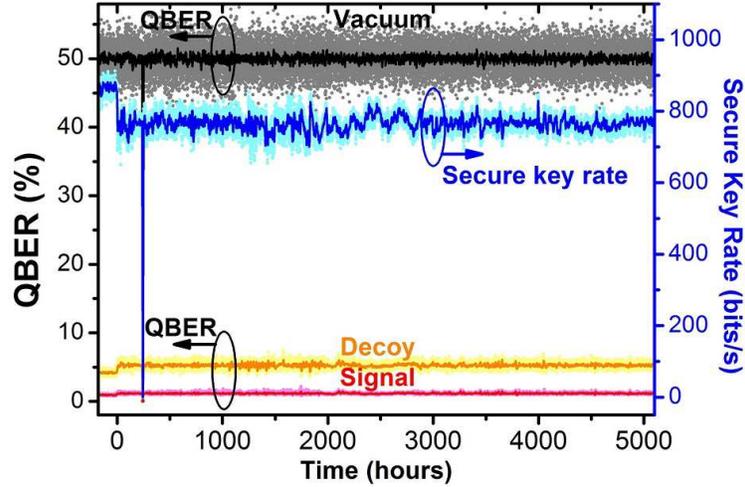} \caption{QBER and
secure key rate varied with time for the Hefei-Chaohu QKD link. }
\label{hc}
\end{figure}

The QBERs of the signal, decoy, and vacuum states were recorded
during the laboratory examination and the field test period (For the
Hefei-Chaohu QKD link see Fig. \ref{hc}, the magenta, yellow, and
dark gray dots stand for the QBERs of the signal, decoy, and vacuum
states, respectively.). And, the secure key rate was also recoded
during the whole period (For the Hefei-Chaohu QKD link see Fig.
\ref{hc}, the cyan dot stands for the secure key rate). In Fig.
\ref{hc}, the laboratory examination stage of the whole QKD network
lasted about one week, started from hour -168.6 and stopped at hour
0. The field test stage of the whole network started from hour 0 and
stopped at hour 5093.9. And there was an interruption lasted for
about 8 hours, because of the power outage in Node WTPT, during
which time the QBERs and secure key rate reduced to zero.

There were obvious changes between the laboratory examination stage
and the field test period for the Hefei-Chaohu QKD link, although
the losses of the fiber channels in this two stage were almost the
same. In the laboratory, the QBERs of the signal state and decoy
state between T5 and R5 were respectively 0.94\% and 4.18\%, and the
average secure key rate was 0.87 kbps. While, in the field
circumstance, the QBERs of the signal state and decoy state
increased up to 1.16\% and 5.26\%, and the secure key rate reduced
to 0.77 kbps. These changes come from the background noise of the
fiber channel and the surrounding environments of the QKD devices T5
and R5. The background noise was mainly the leaky photons from other
parallel fibers in the same multi-core cable. The yield of the
vacuum state was about $4\times10^{-6}$ in the laboratory
examination stage, but increased to $5\times10^{-6}$ in the field
test period. This presented background noise cause about 0.1\%
increase of the QBER of the signal state. The QKD devices T5 and R5
were placed in the telecom rooms of China Mobile Ltd., where the
vibration and acoustic noise were not ignorable. Even though the FMI
was polarization independent and vibration isolation was considered
when we designed QKD devices, the surrounding environments cause
about 0.1\% increase of the QBER of the signal state.

The fluctuations of QBERs and the secure key rate are observed in
Fig. \ref{hc}. We have also noticed that, the ranges of
fluctuations during the field test period was obvious large than the
ones during the laboratory examination stage. In the laboratory,
these fluctuations mainly originated from the statistical
fluctuation and detection probability fluctuation. But, when the QKD
devices were tested in the field, the fluctuations of QBERs and key
rates also come from the variations of fiber loss.

The secure key rates of each QKD links in the network are shown in
table \ref{keyrate}. The performance of the wide area QKD network
was limited by the intercity QKD link, which has lower secure key
rate due to longer channel. For example, in Hefei metropolitan area
QKD network, the NC-WC QKD link has 29.54 kbps secure key rate in
state (\uppercase\expandafter{\romannumeral 3}), while the secure
key rate of Hefei-Chaohu QKD link reduces to 0.77 kbps. We expect to
increase the performance (especially the secure key rate) of the QKD
network in near future with improvements of hardware and software.

\begin{table}[htb]
\centering\caption{Secure key rates of each links in the
Hefei-Chaohu-Wuhu wide area QKD network.} \label{keyrate}
\begin{tabular}{c c c c}
\hline \hline                        &\textbf{QKD link}   &\textbf{QKD devices}   &\textbf{Secure key rate}(kbps)     \\
\hline \textbf{Hefei metro network}  &KLQI$\rightarrow$NC            &T1$\rightarrow$R3      &7.27                                       \\
                                     &NC$\rightarrow$WC              &T3$\rightarrow$R4      &8.33                                         \\
                                     &WC$\rightarrow$KLQI            &T4$\rightarrow$R1      &6.86                                        \\\cline{2-4}
                                     &WTPT$\rightarrow$NC            &T2$\rightarrow$R3      &1.05                                        \\
                                     &NC$\rightarrow$WC              &T3$\rightarrow$R4      &8.33                                           \\
                                     &WC$\rightarrow$WTPT            &T4$\rightarrow$R2      &1.02                                       \\\cline{2-4}
                                     &KLQI$\rightarrow$WTPT          &T1$\rightarrow$R2      &2.67                                        \\
                                     &WTPT$\rightarrow$KLQI          &T2$\rightarrow$R1      &2.42                                           \\
                                     &NC$\rightarrow$WC              &T3$\rightarrow$R4      &13.39                                          \\
                                     &WC$\rightarrow$NC              &T4$\rightarrow$R3      &16.15                                      \\
\hline
\textbf{HCW-intercity link}          &Hefei$\rightarrow$Chaohu       &T5$\rightarrow$R5      &0.77                                      \\
                                     &Chaohu$\rightarrow$Wuhu        &T6$\rightarrow$R6      &0.80                                      \\
\hline
\textbf{Wuhu metro network}          &WHB$\rightarrow$TR             &T7$\rightarrow$R7      &6.07                                             \\
                                     &Qasky$\rightarrow$TR           &T8$\rightarrow$R7      &0.96                                         \\
\hline\hline
\end{tabular}
\end{table}

\section{Performance of the encryption applications}

Just as Stucki et al. said in reference \cite{idq}, the test of the
performance of the encryption application was not the primary goal
of the field QKD network. However, the encryption applications are
the motive force of development of quantum cryptography, and the
integrated technology of QKD device (the quick and secure key
refresh server) and encryptor should be developed. Take the Thales
Mistral products for example, the encryptors could not deal with a
key renewal period lower than 3 seconds \cite{cv}. Here, two typical
applications of the quantum keys from our QKD devices were developed
and tested on this wide area QKD network.

The first one was the one-time pad encryption medium for the public
switch telephone network (PSTN). This encryption medium was added
between PSTN and traditional terminals, such as the conventional
telephone and fax machine. It's very convenient for subscribers who
do not need to change their original telephone and fax machines,
even their operative habits, the encryption medium was completely
transparent in normal circumstance without security requirements.
The keys used for one-time pad encrypted voice data came from the
QKD server in two ways, one was real-time transport, which was
tested on the QCN-USTC, the other one was downloading and storing in
a secure digital (SD) card, which was tested on the link between
KLQI in Hefei and Qasky in Wuhu.

The second one was the symmetric encryption virtual private network
(VPN) security gateway for public networks. This VPN gateway
integrated the QKD technology and the existing classical IPsec
protocol together, and was completely compatible with conventional
hardware and software in the Internet. The 256-bit advanced
encryption standard (AES) encryption was employed to guarantee the
security. The secret seed keys for AES derived from and refreshed by
the QKD server. In this wide area QKD network, the VPN security gate
was tested among subscribers in Hefei and subscribers in Wuhu. The
seed key refresh rate for the VPN gateway was limited by the
intercity QKD link, which could only support the rate about 3
256-bit keys per second. Network applications like the file
transfer, internet telephone, video chat, and multimedia-based
tripartite conference were tested.

\section{Conclusions}

We have shown that the dynamic QKD network can be field installed
and run steadily in wide area for long term. Through sharing
communication infrastructures from China Mobile Ltd., the
Hefei-Chaohu-Wuhu wide area QKD network provided a secure and stable
key distribution platform for subscribes in two cities (Hefei and
Wuhu) that are over 150 kilometers apart. Four QKD nodes in Hefei
city were combined together by two dynamic routing components to
form a typical full-mesh metropolitan core network, in which there
were up to eight different direct QKD links only by multiplexing
four fiber channels. In Wuhu city, the point-to-multipoint
metropolitan network was composed by three QKD nodes and one
1$\times$2 switch to simulate a quantum access network. These two
typical QKD networks and HCW-intercity QKD link ran for more than
5000 hours, and even the QCN-USTC part for about two years, the
stable operation of the Hefei-Chaohu-Wuhu wide area QKD network has
proven the reliability and robustness to the field environment of
QKD networks. We have successfully field tested the wide area QKD
network prototype, and also realized the effective integration
between P2P QKD techniques and networking schemes, owing to the
following developments: (1) Standardized design of QKD devices to
facilitate establishment and maintenance of QKD networks; (2)
Resolution of symmetry problem of QKD devices to clear obstacles in
the way towards cost-effective technology for QKD (especially for
one-way phase coding system), and also convenient to dynamically add
new QKD nodes; (3) Seamless switching in dynamic QKD network based
on stable QKD systems and networking techniques. In short, results
from Hefei-Chaohu-Wuhu wide area QKD network demonstrate that QKD
technology actually has the potential to be widely deployed.

\section*{Acknowledgments}

We thanks China Mobile Ltd. and Network Information Center of
University of Science and Technology of China for providing
communication infrastructures. This work was supported by the
National Natural Science Foundation of China (Grant No.61101137,
No.61201239, No.61205118, and No.11304397), National Basic Research
Program of China (Grants No. 2011CBA00200 and No. 2011CB921200), and
National High Technology Research and Development Program of China
(863 program) (Grant No. 2009AA01A349).

\end{document}